\documentclass{ws-procs9x6}            
\begin{document}
\title{Schwinger Effect in (A)dS and Charged Black Hole}

\author{Sang Pyo Kim}

\address{Department of Physics, Kunsan National University,\\
Kunsan 573-701, Korea\\
$^*$E-mail: sangkim@kunsan.ac.kr}

\begin{abstract}
In an (Anti-) de Sitter space and a charged black hole the Schwinger effect is either enhanced by the Hawking radiation or suppressed by the negative curvature. We use the contour integral method to calculate the production of charged pairs in the global (A)dS space. The charge emission from near-extremal black hole is found from the AdS geometry near the horizon and interpreted as the Schwinger effect in a Rindler space with the surface gravity for the acceleration as well as the Schwinger effect in AdS space.
\end{abstract}


\bodymatter

\section{Introduction}\label{int}

\begin{figure}[b]
\begin{center}
\includegraphics[width=3.0in]{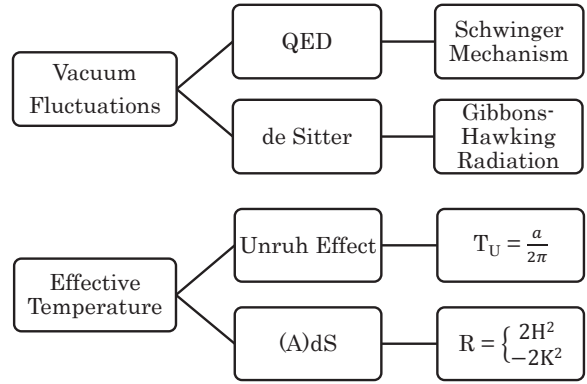}
\end{center}
\caption{The Schwinger effect and the Gibbons-Hawking radiation in dS from quantum fluctuations. The temperature for the Unruh effect in ${\rm (A)dS}_2$ is $T_{\rm eff} = \sqrt{T_{\rm U}^2 + \frac{R}{8 \pi^2}}$.\cite{deser-levin}} \label{fig1}
\end{figure}

A charged Reissner-Nordstr\"{o}m (RN) black hole emits not only the Hawking radiation of all species of
particles due to the horizon but also the Schwinger emission of charged particles due to the electric field
near the horizon, and thus provides an arena where one may explore the intertwinement between quantum
electrodynamics (QED) and quantum aspect of black holes. The pair production by a uniform electric field in an (A)dS space also exhibits the interplay between QED and quantum gravity effect. As summarized in Fig.~\ref{fig1}, vacuum fluctuations produce pairs from the vacuum and the horizon of a black hole or dS space and the electric field physically separate pairs. The black hole radiation has a thermal distribution with the Hawking temperature as a Unruh temperature with the surface gravity on the horizon.\cite{unruh} Cai and Kim show that the Schwinger effect in the (A)dS space has also a thermal interpretation with a Unruh temperature of accelerating charges in that space.\cite{cai-kim14} Using the geometry ${\rm AdS}_2 \times {\rm S}^2$ near the horizon of a near-extremal black hole, Chen et al have explicitly found the Schwinger effect from the charged black hole.\cite{cklsw,cstt}

In this paper, after briefly reviewing the recently introduced contour integral method,\cite{kim13a} we apply it to
the pair production by a uniform electric field in the global ${\rm (A)dS}_2$ geometry.
Then, using the near-horizon geometry ${\rm AdS}_2 \times {\rm S}^2$ of a near-extremal black hole,
we find the Schwinger effect from the charged black hole
and propose a thermal interpretation of the charge emission.\cite{kim-lee-yoon15,kim15}
In the in-in formalism, the quantum field theory is extended to the complex plane of time or space
and the out-vacuum transported along a loop starting from the past infinity is distinguished from the
in-vacuum by simple poles of the frequency or momentum in the background
spacetime and/or electromagnetic field.\cite{kim13a} Hence, particle production is characterized and determined
by polons, that is, contour integrals of all homotopy classes of winding number one encircling poles.\cite{kim13a,kim13b,kim14a} Differently from the planar coordinates without any finite pole, a massive scalar field in the global coordinates has simple poles both at the north and south pole of the global geometry, whose contributions interfere and result in the Stokes phenomenon of the dS radiation depending on the spacetime dimension.\cite{kim13b} It is investigated whether the Schwinger effect may have such a Stokes phenomenon in the global ${\rm dS}_2$ geometry. 
The holographic Schwinger effect is also studied in the dS space\cite{fnpt} and AdS space.\cite{chen15}

\section{Pair Production via Contour Integrals}\label{sec2}

In the functional Schr\"{o}dinger picture and the in-in formalism, the in-vacuum in the past infinity evolves to the future infinity and return to the past infinity, which gives the mean number of pairs\cite{kim13a}
\begin{eqnarray}
{\cal N}_{\kappa} = \Bigl| \sum_{J} \langle 0_{\kappa}, {\rm in} \vert 0_{\kappa}, C^{(1)}_J \rangle^2 \Bigr|
=  \Bigl| \sum_{J} e^{- i \oint_{C^{(1)}_J } \omega_{\kappa} (z) dz} \Bigr|. \label{con int}
\end{eqnarray}
Here, the quantum evolution is extended to the complex plane and $C^{(1)}_J$ denotes distinct contours of winding number one.
For instance, the global dS geometry has simple poles not only at infinity but also at the north and south poles, which contribute constructively or destructively depending on whether the spacetime dimension is even or odd.\cite{kim13b} On the other hand, the planar dS space has the simple pole at infinity only and has a nondestructive radiation with the Gibbons-Hawking temperature. A physical reasoning may be that the propagator and hence the Hyugens principle exhibits a different nature. In the in-in formalism, the scattering matrix between the in-vacuum and another in-vacua tranported along contours $C^{(n)}_J$ of all winding numbers gives the vacuum persistence
\begin{eqnarray}
2 {\rm Im} {\cal L}_{\rm eff} =  \pm \sum_{\kappa}  \Bigl| \sum_{J} \sum_{n} \ln (\langle 0_{\kappa}, {\rm in} \vert 0_{\kappa}, C^{(n)}_J \rangle^2) \Bigr| = \pm \sum_{\kappa} \ln (1 \pm {\cal N}_{\kappa}), \label{vac per}
\end{eqnarray}
where the upper (lower) sign is for scalar (spinor) QED and the mean number ${\cal N}_{\kappa}$ is determined by the contours of winding number one in Eq. (\ref{con int}).

\section{Schwinger Effect in (A)dS}\label{sec3}

The production of charged pairs by a uniform electric field in a dS space exhibits both the Schwinger effect and the Gibbons-Hawking radiation as shown in Fig.~\ref{fig1}.\cite{garriga,kim-page08} An AdS space confines virtual pairs and limits the Schwinger pair production through the Breitenlohner-Friedman bound.\cite{pioline-troost,kim-page08,cai-kim14}
Now, we consider the Schwinger effect in the two-dimensional global dS geometry
\begin{eqnarray}
ds^2 = - dt^2 +  \cosh^2 (Ht) dx^2,
\end{eqnarray}
and in the uniform electric field with a vector potential $A_1 = - E \sinh (Ht)/H$. We distinguish the weak-field limit ($qE \ll H^2/2$) from the strong-field limit ($qE \gg H^2/2$), in which the pair production has different characteristics.

First, we consider the weak-field limit ($\gamma > 0, \lambda > 0$ below). Following Ref.~\citenum{kim13b}, a charged scalar
has the time-dependent frequency
\begin{eqnarray}
\omega^2 (t) = (\gamma H)^2 + \frac{(\lambda H)^2}{\cosh^2 (Ht)}, \label{freq}
\end{eqnarray}
where
\begin{eqnarray}
\gamma = \sqrt{ \Bigl( \frac{qE}{H^2} \Bigr)^2 + \frac{m^2}{H^2}  - \frac{1}{4}}, \quad \lambda = \sqrt{\frac{1}{4} - \Bigl( \frac{qE}{H^2} \Bigr)^2}.
\end{eqnarray}
The frequency (\ref{freq}) has two poles at $\tau = \pm i \pi/(2H)$ corresponding to the north and south pole of the global geometry, which lead to
the so-called Stokes phenomenon of dS radiation.\cite{kim13b} As will be shown below, the Schwinger effect has also the Stokes phenomenon in the weak-field limit.
Let us introduce a conformal transformation $z = e^{Ht}$ and restrict to a single Riemann sheet. Then, the contour integral takes the form
\begin{eqnarray}
\oint \omega(t) dt = \gamma \oint \frac{dz}{z(z^2+1)} \sqrt{(z-z_1)(z-z_1^*)(z-z_2)(z-z_2^*)}, \label{ds freq}
\end{eqnarray}
where
\begin{eqnarray}
z_1 = i \Bigl(\sqrt{1+ \frac{\lambda^2}{\gamma^2}} + \frac{\lambda}{\gamma} \Bigr), \quad
z_2 = i \Bigl(\sqrt{1+ \frac{\lambda^2}{\gamma^2}} - \frac{\lambda}{\gamma}\Bigr).
\end{eqnarray}
The integrand has finite simple poles at $z = \pm i$ corresponding to the north and south pole of the global geometry and another simple pole from $z = \infty$. The residues at $z = \pm i$ take the opposite signs due to the branch cuts. Summing over four distinct contours of winding number one, we obtain from Eq. (\ref{con int})
\begin{eqnarray}
{\cal N}_{\rm dS} = \bigl(e^{2 i \pi \lambda} + 2 + e^{-2 i \pi \lambda} \bigr) e^{- 2 \pi \gamma} = 4 \cos^2 (\pi \lambda) e^{- 2 \pi \gamma}. \label{ds weak}
\end{eqnarray}
The oscillatory factor is the characteristic of the Stokes phenomenon.
\begin{figure}[t]
\begin{center}
\includegraphics[width=3.0in]{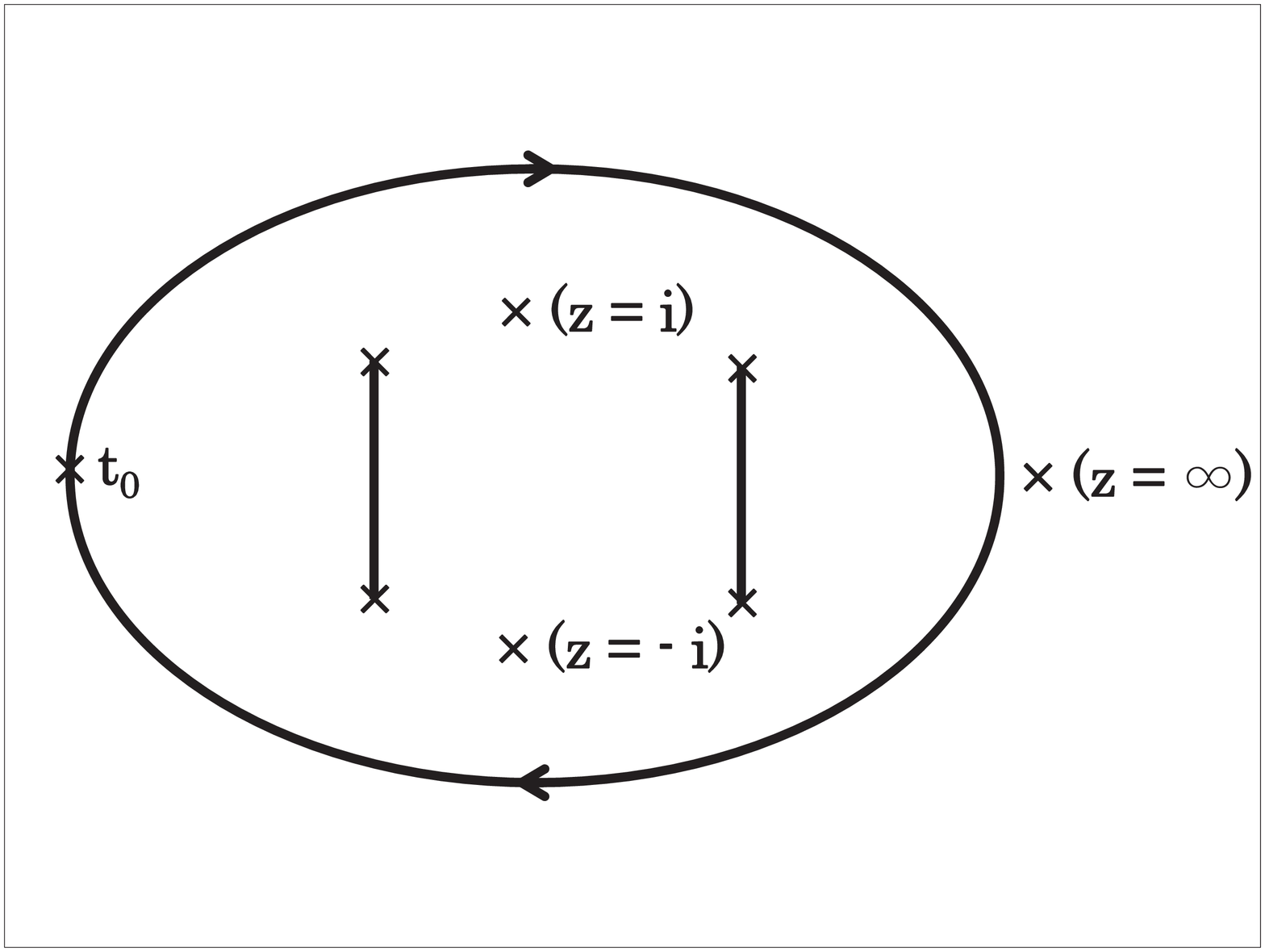}
\end{center}
\caption{The frequency (\ref{ds freq}) for a charged scalar in a uniform electric field and the global ${\rm dS}_2$ has two pairs of branch cuts in the complex plane and a pair of simple poles at $z= \pm i$ and another simple pole at $z= \infty$. The contour includes both poles $z= \pm i$, whose residues sum vanishes but receives a contribution from $z= \infty$.}\label{fig2}
\end{figure}

Second, in the strong field limit ($\gamma > 0, \bar{\lambda} > 0$) the integrand has different roots
\begin{eqnarray}
z_{1}^{(\pm)} = \frac{\bar{\lambda}}{\gamma} \pm i \sqrt{1- \frac{\bar{\lambda}^2}{\gamma^2}}, \quad z_{2}^{(\pm)} = - \frac{\bar{\lambda}}{\gamma} \pm i \sqrt{1- \frac{\bar{\lambda}^2}{\gamma^2}},
\end{eqnarray}
where
\begin{eqnarray}
\bar{\lambda} = \sqrt{\Bigl( \frac{qE}{H^2} \Bigr)^2 -\frac{1}{4}}.
\end{eqnarray}
Then, the mean number of pairs from four distinct contours, for instance Fig.~\ref{fig2}, is given by
\begin{eqnarray}
{\cal N}_{\rm dS} = \bigl(e^{2 \pi \bar{\lambda}} + 2 + e^{-2 \pi \bar{\lambda}} \bigr) e^{- 2 \pi \gamma} = 4 \cosh^2 (\pi \bar{\lambda}) e^{- 2 \pi \gamma}. \label{ds strong}
\end{eqnarray}
The mean number (\ref{ds strong}) is consistent with the exact result from the Bogoliubov transformation using the exact solution.\cite{kim-hwang-wang}

Finally, a charged scalar in the global AdS geometry
\begin{eqnarray}
ds^2 = - \cosh^2 (Kx) dt^2 + dx^2,
\end{eqnarray}
and in a vector potential $A_0 = - E \sinh (Kx)/K$ for the uniform electric field has the momentum
\begin{eqnarray}
q^2 (x) = (\gamma_{\rm AdS} K)^2 - \frac{(\lambda_{\rm AdS} K)^2}{\cosh^2 (Kx)},
\end{eqnarray}
where
\begin{eqnarray}
\gamma_{\rm AdS} = \sqrt{\Bigl( \frac{qE}{K^2} \Bigr)^2 - \frac{m^2}{K^2} - \frac{1}{4}}, \quad \lambda_{\rm AdS} = \sqrt{\Bigl( \frac{qE}{K^2} \Bigr)^2 +\frac{1}{4}}.
\end{eqnarray}
As for the dS space, four distinct contours contribute to the mean number
\begin{eqnarray}
{\cal N}_{\rm AdS} = 4 \sinh^2 (\pi \lambda_{\rm AdS}) e^{- 2 \pi \gamma_{\rm AdS}}. \label{ads strong}
\end{eqnarray}

\section{Schwinger Effect in Near-Extremal Charged Black Hole}\label{sec4}

The near-extremal black hole has the near-horizon geometry\cite{cklsw,cstt}
\begin{eqnarray}
ds^2 = - \frac{\rho^2 - B^2}{Q^2} d\tau^2 + \frac{Q^2}{\rho^2 - B^2} d \rho^2 + Q^2 d \Omega_2^2, \label{bh ads}
\end{eqnarray}
where the coordinates are stretched as $\rho= (r - Q)/\epsilon, \tau= \epsilon t$ and the deviation of the charge from the mass is measured as $M - Q = (\epsilon B)^2/(2Q)$. The Coulomb potential is $A_0 = - \rho/Q$ on the horizon.
The geometry (\ref{bh ads}) is the planar coordinates of ${\rm AdS}_2 \times {\rm S}^2$. The radial momentum for the spherical harmonic $Y_{lm}$ and energy $\omega$ is given by
\begin{eqnarray}
S (\rho) = \int \frac{\sqrt{(q \rho - \omega Q)^2 Q^2 - \Bigl(m^2 Q^2 + (l+ 1/2)^2 \Bigr) (\rho^2 - B^2)}}{\rho^2 - B^2} d \rho. \label{HJ action}
\end{eqnarray}
Note that pairs are produced when the BF bound $(Q < l/q)$ for ${\rm AdS}$ is violated.\cite{cai-kim14} Taking a contour of figure eight counterclockwise enclosing $\rho = B$ and $\rho = -B$ gives the leading term ${\cal N} = e^{- ({\cal S}_{a} - {\cal S}_{b})}$, whereas the exact formula from the field equation is given by\cite{cklsw,cstt}
\begin{eqnarray}
{\cal N} = \Biggl( \frac{e^{- ({\cal S}_{a} - {\cal S}_{b})} - e^{- ({\cal S}_{a} + {\cal S}_{b}) }}{1 + e^{- ({\cal S}_{a} + {\cal S}_{b}) }} \Biggr) \Biggl( \frac{1 -  e^{- ({\cal S}_{c} - {\cal S}_{a}) }}{1 + e^{- ({\cal S}_{c} - {\cal S}_{b}) }} \Biggr),
\end{eqnarray}
where
\begin{eqnarray}
{\cal S}_{a} = 2 \pi q Q, \, {\cal S}_{b} = 2 \pi qQ \sqrt{1 - \Bigl(\frac{m}{q} \Bigr)^2 - \Bigl(\frac{l+ 1/2}{qQ} \Bigr)^2 }, \, {\cal S}_c = 2 \pi \frac{\omega}{B} Q^2.
\end{eqnarray}

Interestingly, the mean number of emitted particles consists of three parts, the Boltzmann amplification factor, the Schwinger effect in ${\rm AdS}_2$, and the Schwinger effect in a two-dimensional Rindler space:\cite{kim-lee-yoon15,kim15}
\begin{eqnarray}
{\cal N} = e^{\frac{\bar{m}}{T_{\rm CK}}} \times \Biggl\{ \frac{e^{- \frac{\bar{m}}{T_{\rm CK}}} - e^{- \frac{\bar{m}}{\bar{T}_{\rm CK}}}}{1 + e^{- \frac{\bar{m}}{\bar{T}_{\rm CK}}}} \Biggr\} \times \Biggl\{ \frac{e^{- \frac{\bar{m}}{T_{\rm CK}}} \Bigl(1 - e^{- \frac{\omega - q A(r_{\rm H})}{T_{\rm H}}} \Bigr)}{1+ e^{- \frac{\omega - q A(r_{\rm H})}{T_{\rm H}}} e^{- \frac{\bar{m}}{T_{\rm CK}}}} \Biggr\}. \label{sch-haw}
\end{eqnarray}
Here, $\bar{m} = \sqrt{m^2 + (l+1/2)^2/Q^2}$, $T_{\rm H}$ is the Hawking temperature, and $T_{\rm CK}$ and $\bar{T}_{\rm CK}$ are the effective temperature in ${\rm AdS}_2$ introduced by Cai and Kim\cite{cai-kim14} 
\begin{eqnarray}
T_{\rm CK} = T_{\rm U} + \sqrt{T_{\rm U}^2 - \Bigl( \frac{1}{2 \pi Q} \Bigr)^2}, \quad \bar{T}_{\rm CK} = T_{\rm U} - \sqrt{T_{\rm U}^2 - \Bigl( \frac{1}{2 \pi Q} \Bigr)^2}, \label{eff tem}
\end{eqnarray}
where $T_{\rm U} = q/(2 \pi \bar{m} Q)$ is the Unruh temperature for charge accelerated by the electric field on the horizon.
The Schwinger effect and vacuum persistence in the Rindler space has been first calculated by Gabriel and Spindel.\cite{gabriel-spindel}

\section*{Acknowledgments}
The author thanks Rong-Gen Cai and Chiang-Mei Chen for useful discussions. The participation of ICGAC12 was supported in part by APCTP.
This work was supported by IBS (Institute for Basic Science) under IBS-R012-D1.


\begin{thebibliography}{10}

\bibitem{unruh} W.~G.~Unruh, Notes on black-hole evaporation, {\em Phys. Rev. D} {\bf 14}, 870 (1976).

\bibitem{cai-kim14} R-G.~Cai and S.~P.~Kim, One-loop effective action and Schwinger effect in (anti-) de Sitter space, {\em JHEP} {\bf 09} (2014) 72.

\bibitem{cklsw} C-M.~Chen, S.~P.~Kim, I-C.~Lin, J-R.~Sun and M-F.~Wu, Spontaneous pair production in Reissner-Nordström black holes, {\em Phys. Rev. D} {\bf 85},  124041 (2012).

\bibitem{cstt} C-M.~Chen, J-R.~Sun, F-Y.~Tang and P-Y.~Tsai, Spinor particle creation in near extremal Reissner--Nordström black holes, {\em Class. Quantum Grav.} {\bf 32}, 195003 (2015).

\bibitem{kim13a} S.~P.~Kim, New geometric transition as origin of particle production in time-dependent backgrounds, {\em Phys. Lett. B} {\bf 725},  500 (2013).

\bibitem{kim-lee-yoon15} S.~P.~Kim, H.~K.~Lee, and Y.~Yoon, Thermal Interpretation of Schwinger Effect in Near-Extremal
Reissner-Nordstr\"{o}m Black Hole [arXiv:1503.00218].

\bibitem{kim15} S.~P.~Kim, Schwinger Effect, Hawking Radiation and Gauge-Gravity Relation [arXiv:1506.03990].

\bibitem{kim13b} S.~P.~Kim, Geometric origin of Stokes phenomenon for de Sitter radiation, {\em Phys. Rev. D} {\bf 88},  044027 (2013).

\bibitem{kim14a} S.~P.~Kim, Particle Production from Geometric Transition in Expanding Universe, {\em JPS Conf. Proc.} {\bf 1},  013110 (2014).

\bibitem{fnpt} W.~Fischler, P.~H.~Nguyen, J.~F.~Pedraza, and W.~Tangarife, Holographic Schwinger effect in de Sitter space, {\em Phys. Rev. D}  {\bf 91}, 086015 (2015).
    
\bibitem{chen15} C-M.~Chen, Pair Production in Near Extremal Charged Black Holes, in this Proceedings (2015).  

\bibitem{deser-levin} S. Deser and O. Levin, Accelerated detectors and temperature in (anti-) de Sitter spaces, {\em Class. Quantum Grav.} {\bf 14}, L163 (1997).

\bibitem{garriga} J.~Garriga,  Pair production by an electric field in (1+1)-dimensional de Sitter space, {\em Phys. Rev. D} {\bf 49}, 6343 (1994).

\bibitem{kim-page08} S.~P.~Kim and D.~N.~Page, Schwinger pair production in ${\rm dS_2}$ and ${\rm AdS_2}$, {\em Phys. Rev. D} {\bf  78}, 103517  (2008).

\bibitem{pioline-troost} B.~Pioline and J.~Troost, {\em JHEP} {\bf 03} (2005) 043.    

\bibitem{kim-hwang-wang} S.~P.~Kim, P.~W-Y.~Hwang and T-C.~Wang, Schwinger mechanism in ${\rm dS_2}$ and ${\rm AdS_2}$ revisited [arXiv:1112.0885].

\bibitem{gabriel-spindel} Cl.~Gabriel and Ph.~Spindel, Quantum Charged Fields in (1+1) Rindler Space, {\em Ann. Phys.} {\bf 284}, 263 (2000).

\end{thebibliography}
\end{document}